\documentclass[notitlepage,amsmath,amssymb,aps,reprint,superscriptaddress]{revtex4-1} 
\usepackage{graphicx}
\usepackage{dcolumn}
\usepackage{bm}
\usepackage{braket}
\usepackage{amsmath}
\usepackage{bm}
\usepackage{amssymb}
\usepackage{setspace}
\usepackage[euler]{textgreek}
\allowdisplaybreaks
\usepackage{multirow}
\usepackage{comment}
\usepackage{relsize}
\usepackage{hyperref}
\usepackage{siunitx}
\hypersetup{
    colorlinks=true,
    linkcolor=blue,
    filecolor=magenta,      
    urlcolor=cyan,
    citecolor=red
}

\newcommand{\etalink}{\eta_{\text{link}}}
\newcommand{\etadev}{\eta_{\text{PDR}}}
\newcommand{\etadet}{\eta_{\text{det}}}
\newcommand{\etapol}{\eta_{\text{pol}}}
\newcommand{\etapolV}{\eta_{\text{pol,V}}}
\newcommand{\etapolH}{\eta_{\text{pol,H}}}
\newcommand{\rcav}{R_{\text{cav}}}
\newcommand{\rcavV}{R_{\text{cav,V}}}
\newcommand{\rcavH}{R_{\text{cav,H}}}
\newcommand{\pdet}{p_{\text{det}}}

\newcommand{\plost}{p_{\text{lost}}}
\newcommand{\tpulse}{\tau_{\text{pulse}}}
\newcommand{\treset}{\tau_{\text{reset}}}

\newcommand{\Nmax}{N_{\text{max}}}

\begin{document}
\title{A Polarization Encoded Photon-to-Spin Interface}
\date{\today}
\author{K. C. Chen}
\author{E. Bersin}
\author{D. Englund}
\affiliation{Research Laboratory of Electronics, Massachusetts Institute of Technology, Cambridge, Massachusetts 02139, USA}
\begin{abstract}
We propose an integrated photonics device for mapping qubits encoded in the polarization of a photon onto the spin state of a solid-state defect coupled to a photonic crystal cavity: a `Polarization-Encoded Photon-to-Spin Interface' (PEPSI). We perform a theoretical analysis of the state fidelity's dependence on the device's polarization extinction ratio and atom-cavity cooperativity. Furthermore, we explore the rate-fidelity trade-off through analytical and numerical models. In simulation, we show that our design enables efficient, high fidelity photon-to-spin mapping.
\end{abstract}
\maketitle

\section{Introduction}

Quantum networks are being developed to distribute entanglement across distant nodes. A central requirement common to most quantum network approaches is the development of efficient interfaces between spins and photonic qubits~\cite{Wehner_2018}. Among the various photonic degrees of freedom, encoding in the polarization basis \{$\ket{H},\ket{V}$\} (horizontal and vertical) is attractive over number-state encoding because photon loss becomes a heralded error~\cite{Barrett_2005}.

To couple the polarization-encoded photon qubit $\ket{\psi_P}=\alpha\ket{H} +\beta\ket{V}$ to an atomic memory in a network node, Duan and Kimble~\cite{Duan_2004} proposed the scheme illustrated in Fig.~\ref{fig:concept}(a). The incoming photon passes through a polarizing beam splitter (PBS) so that only the $V$ polarization is reflected off a single sided cavity whose mode couples with the $|\downarrow\rangle\longleftrightarrow|\downarrow'\rangle$ transition. The $H$ polarization is reflected off a mirror, and recombines with the $V$ polarization to form an entangled spin-photon state: $\ket{\psi_{\text{ent,out}}}=-\alpha\ket{H,\downarrow}+\beta\ket{V,\downarrow}-\alpha\ket{H,\uparrow}-\beta\ket{V,\uparrow}$. Subsequent measurement of the photonic state heralds the transfer of the polarization qubit to the atom, as demonstrated in recent experiments using trapped neutral atoms~\cite{Tiecke_2014} and diamond color centers~\cite{Bhaskar_2019}.

However, in a free-space setup, a major technical challenge concerns the need to maintain stability of the phase difference between two spatially separated meters-long polarization paths~\cite{Tiecke_2014}. In this Article, we propose a monolithic, micron-scale photonic structure that combines the $H$ and $V$ paths into one phase-stable architecture (Fig.~\ref{fig:concept}(b)). We estimate that this system will enable state transfer fidelity exceeding $99\%$. This Polarization-Encoded Photon-to-Spin Interface (PEPSI) greatly simplifies quantum networking with polarization-encoded photons coupled to atomic memories.

\section{Results} \label{monolithic}

\subsection{Device} \label{sec_device}

\begin{figure}
    \centering
    \includegraphics[width=0.5\textwidth]{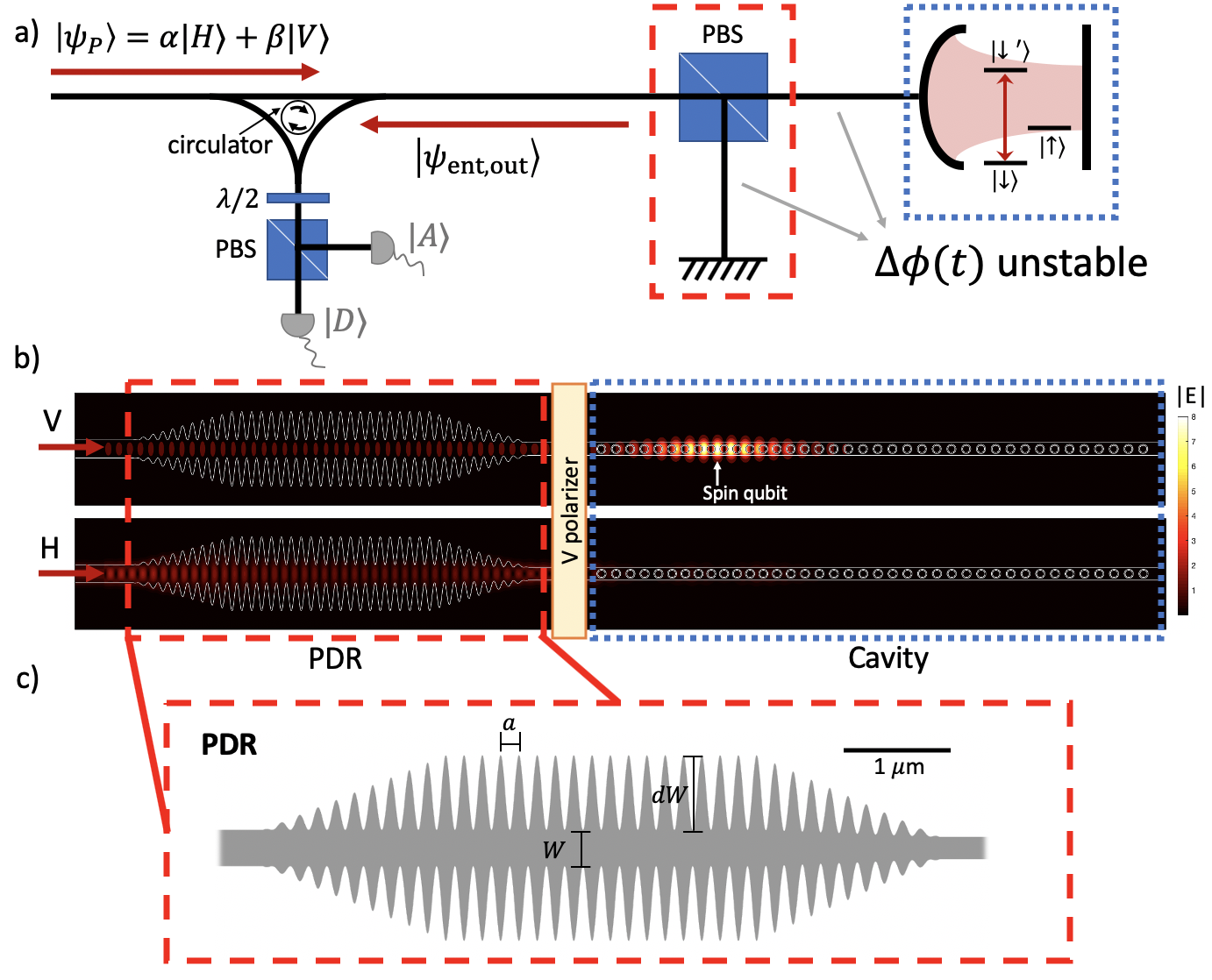}
    \caption{(a) The Duan-Kimble scheme for polarization-spin mapping. The requirements of a high extinction polarizing beam splitter (PBS), strong cavity-atom coupling, and stabilization of the phase mismatch between arms $\Delta\phi(t)$ all make implementation with bulk optics challenging.  (b) Our proposal for a phase-stable monolithic device (PEPSI) that implements the protocol in (a). A polarization-dependent reflector (PDR, red dashed lines) on the input partially reflects $H$ light while passing $V$ light through to interact with the cavity-emitter system (blue dashed lines). A $V$-pass polarizer after the PDR to suppress any undesired atom excitation by the transmitted $H$ light. (c) A zoom-in depiction of the PDR with geometry parameters $a$ (periodicity), $W$ (width), and $dW$ (modulation amplitude).}
    \label{fig:concept}
\end{figure}

Qubits based on single atoms such as solid state color centers, neutral atoms, and trapped ions have emerged as promising systems for quantum networking applications. In particular, group-IV color centers in diamond are attractive candidates due to their excellent optical and spin coherence~\cite{Sukachev_2017,Pingault_2017,Siyushev_2017,Trusheim_2020}. In this Article, we focus on the negatively-charged silicon-vacancy center (SiV) coupled to a diamond nanocavity, though the approach generalizes to other stationary qubits such as neutral atoms~\cite{Tiecke_2014}, trapped ions~\cite{Stute_2012,Takahashi_2020}, and quantum dots~\cite{Sun_2016,Sun_2018}.

\begin{figure*}
    \centering
    \includegraphics[width=\textwidth]{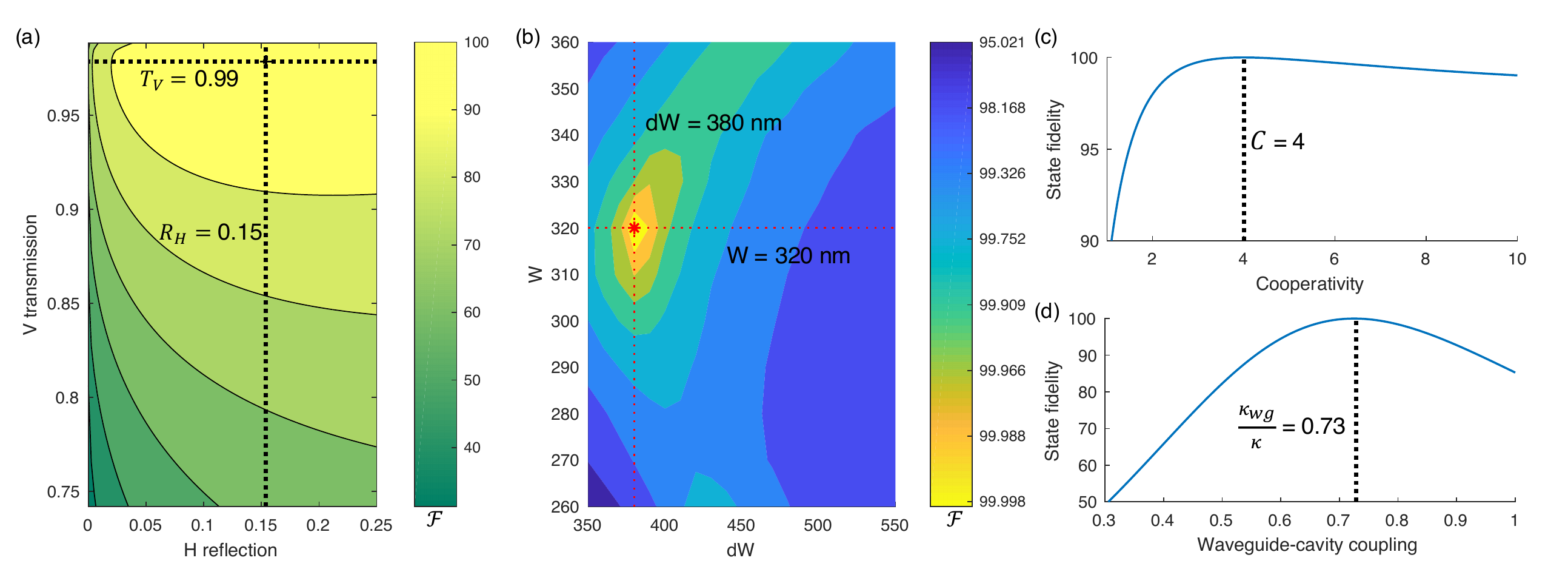}
    \caption{Fidelity as a function of PDR parameters, atom-cavity cooperativity $C$, and waveguide-cavity coupling $\kappa_{wg}/\kappa$. (a) The state fidelity is plotted as a function of PDR $T_V$ ($V$ transmission) and $R_H$ ($H$ reflection). The maximum $\mathcal{F}$ occurs when $T_V=0.99$ and $R_H=0.15$. (b) $\mathcal{F}$ is plotted as a function of the two PDR geometry parameters: the width $W$ and the modulation amplitude $dW$. $W=320~$nm and $dW=380~$nm are chosen for our particular device with an optimum $\mathcal{F}=99.996\%$. (c,d) The state fidelity as functions of the atom-cavity cooperativity and waveguide-cavity coupling. The PDR is designed specifically for a single sided cavity with $C=4$ and $\kappa_{wg}/\kappa=0.73$ such that any deviations would introduce loss imbalance that degrades the fidelity.}
    \label{fig:imperfect}
\end{figure*}

As illustrated in Fig.~\ref{fig:concept}(b), the structure comprises i)~a polarization-dependent reflector (PDR) for the $H$~(TE) mode, ii)~a $V$-pass polarizer (see Supplemental Information), and iii)~an over-coupled single sided cavity for the $V$~(TM) mode. The PEPSI collapses both interferometric arms into one co-propagating path that greatly suppresses phase instability stemming from environmental noise. In contrast, bulk optics suffer from thermal and vibrational fluctuations that incur phase noise, which requires phase stabilization costly in operation time\cite{Bhaskar_2019}.

The PDR shown in Fig.~\ref{fig:concept}(c) uses a corrugated photonic crystal design with periodicity $a=169~$nm, width $W=1.83a$, modulation amplitude $dW=2.9a$. An adiabatic taper transfers photons to a 1D photonic crystal nanocavity coupled to the SiV. The remainder of this Article analyzes the performance this phase-stable device. Specifically, we investigate the impact of PEPSI device parameters on state transfer fidelity (Sec.~\ref{sec_imperfect}), the rate-fidelity trade-off in a quantum network link (Sec.~\ref{sec_rate}), and extensions to a scalable photonic integrated circuit platform (Sec.~\ref{integrated}).

\subsection{Effects of device imperfections} \label{sec_imperfect}

To analyze the state transfer process, we consider a single photon ($\ket{\psi_P}$) incident on a cavity-coupled spin qubit $\ket{\psi_{s,i}}=(|\uparrow\rangle+|\downarrow\rangle)/\sqrt{2}$. Using a Schr\"{o}dinger picture evolution, we calculate the resulting spin state $\ket{\psi_{s,f}(i)}$ after detection of the reflected photon in the polarization diagonal basis \{$\ket{D},\ket{A}$\}. Its overlap with the desired state transferred from the polarization qubit defines the state fidelity $\mathcal{F}$, of which we take the average~\cite{Chuang_2010}:
\begin{align}
\mathcal{F} &= \frac{1}{4}\sum_i\mathcal{F}_i = \frac{1}{4}\sum_i|\langle\phi_i|\psi_{s,f}(i)\rangle|^2
\label{eq:fid_single}
\end{align}
where $\ket{\phi_{1,2}}=(|\downarrow\rangle\pm |\uparrow\rangle)/\sqrt{2}$, $\ket{\phi_{3,4}}=(|\downarrow\rangle\pm i|\uparrow\rangle)/\sqrt{2}$.

When a device is perfect such that a lossless PDR has infinite polarization extinction ratios and a nanocavity has perfect waveguide-cavity coupling  ($\kappa_{wg}/\kappa=1$), the cavity reflection solely determines the fidelity that scales as $(C-1)/(C+1)$ in the large cooperativity limit~\cite{Tiecke_2014}. However, when the PDR has finite extinction ratios and scattering losses, and the single sided cavity has a reduced waveguide-cavity coupling efficiency $\kappa_{wg}/\kappa<1$, the need to balance losses becomes especially important to achieving high fidelity. For example, considering the desired state $\ket{\phi_1}$ where $\alpha=\beta=1$, balancing losses entails matching the two coefficients $|r_{H,on}-r_{V,on}|=|r_{H,off}+r_{V,off}|$, which are both functions of PDR transmissivity/reflectivity and the cavity reflectivity (see Methods). Fig.~\ref{fig:imperfect}(a) shows $\mathcal{F}$ as a function of the PDR's $V$ transmission ($T_V$) and $H$ reflection ($R_H$) coefficients given a low scattering loss and fixed cavity parameters corresponding to our design: waveguide-cavity coupling $\kappa_{wg}/\kappa\approx 0.73$ and cooperativity $C\approx 4$ (see Supplemental Information). We find the fidelity is maximized at $99.996\%$ (assuming perfect gate and detection fidelities) when $T_V=0.99$ and $R_H=0.15$, corresponding to transmission and reflection extinction ratios of $0.68~$dB and $12.3~$dB for 20 periodicities. Additionally, we analyze the fidelity's dependence on the PDR geometry by sweeping both $W$ and $dW$. Figure~\ref{fig:imperfect}(b) indicates that a PDR with $W=320~$nm and $dW=380~$nm satisfy both $T_V\approx 0.99$ and $R_H\approx 0.15$. We calculate that $\mathcal{F}$ still well exceeds $99\%$ for any $dW$ between $\sim$350~nm and $\sim$450~nm and similarly for any $W$ between $\sim$280~nm and $\sim$350~nm, providing the PEPSI tolerance to fabrication errors.

Since the PDR design intimately relies on the cavity parameters, we show that the state fidelity does not improve by naively increasing the atom-cavity cooperativity and waveguide-cavity coupling. Fig.~\ref{fig:imperfect}(c) shows that $\mathcal{F}$ reaches its maximum when the cooperativity $C$ approaches 4. Any further increase in $C$ begins to lower the fidelity due to an increasing loss imbalance. Likewise, Fig.~\ref{fig:imperfect}(d) shows that $\mathcal{F}$ is maximized when waveguide-cavity coupling is $\kappa_{wg}/\kappa=0.73$. An under-coupled or critically coupled cavity would result in a severely degraded state transfer with fidelity below $80\%$ due to insufficient cavity reflection. On the other hand, over-coupling past $\kappa_{wg}/\kappa=0.73$ would curtail the atom-cavity interaction and consequently lowers the fidelity.

\subsection{Quantum state transfer rate} \label{sec_rate}

\begin{figure}
    \centering
    \includegraphics[width=0.45\textwidth]{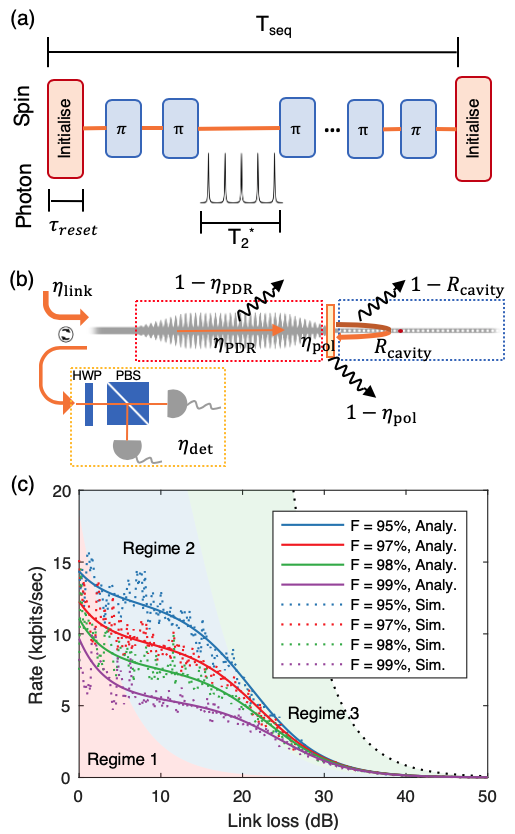}
    \caption{(a) Pulse sequences for conducting quantum state transfer between a polarization qubit and a spin qubit. (b) A diagram depicting where scattering losses occur. (c) Transfer rate (kilo-qubits per second, operated at 5.81~MHz clock rate) as a function of link loss $1-\eta_{\text{link}}$ for four fidelity constraints: $\mathcal{F}=95,97,98,99\%$. We plot both the analytical solutions (solid) and the Monte Carlo simulations (dashed). We categorize the rate as a function of link loss into Regime 1 (red), Regime 2 (blue), and Regime 3 (green). The black dashed line denotes the repeaterless bound for quantum key distribution protocols~\cite{Pirandola_2017}.}
    \label{fig:rate}
\end{figure}

We now analyze the performance of the PEPSI in facilitating quantum state transfer across a lossy network link by looking at the rate-fidelity trade-off for a device realized in simulation, parameterized in Sec.~\ref{sec_device}. We denote the probability of a photon entering and returning through the PDR as $\etadev^2\etapol^2\rcav$, where $\rcav$ is the cavity reflection coefficient and $1-\rcav$ is the cavity decay rate into the environment (assume negligible transmission through the single sided cavity). $\etadev$ and $\etapol$ are the transmission efficiencies of the PDR and the polarizer. As shown in Figure \ref{fig:rate}(a), the protocol begins by initializing the spin qubit in a superposition state $(\ket{\uparrow}+\ket{\downarrow})/\sqrt{2}$ in a time $\treset=\SI{30}{\micro\second}$ as demonstrated in Ref.~\cite{Bhaskar_2019}. A photonic qubit $\ket{\psi_P}$ launched across the link with transmissivity $\etalink$ reaches the PDR shown in Fig~\ref{fig:rate}(b). If the reflected photon is detected as described above with probability $\pdet=\etalink\etadev^2\etapol^2\rcav\etadet$, the spin qubit is projected to the state $\alpha\ket{\downarrow}+\beta\ket{\uparrow}$ (see Methods) with fidelity given by Eq.~\ref{eq:fid_single}. If no photon is detected, the protocol is repeated. 

However, when $\etalink\ll 1$, most transmission attempts do not interact with the spin, and time can be saved by not re-initializing on every transmission attempt. In particular, we consider a series of $N$ photons injected into the link after spin re-initialization. These photons are interspersed by dynamical decoupling pulses ($\pi$-pulses in Fig.~\ref{fig:rate}(a)) to maintain memory coherence. If the detector registers a click for the $m$th attempt, the receiver blocks the subsequent $N-m$ pulses. The complication is that any photon that reaches the cavity but is subsequently lost produces an unheralded error with probability $p_e=\etalink\etadev\etapol(\rcav(1-\etapol+\etapol(1-\etadev))+1-\rcav)$, since the environment projects the spin to a mixed state $\rho_{\text{mixed}}=\frac{1}{2}\mathbb{I}$. Thus, the optimum fidelity for a given device is achieved by re-initializing the spin qubit in advance of every photon transmission. However, the photon can also be either lost in the link before reaching the spin with probability $\plost=1-p_e-\pdet$ or heralded by the detector with probability $\pdet$. Conditioned on not detecting a click, the probability of photon loss without contaminating the spin is $\plost/(1-\pdet)$, and therefore the probability of photon never reaching the spin after $m-1\leq N$ channel uses is $(\plost/(1-\pdet))^{m-1}$.

Given a detector click on the $m^{\text{th}}$ attempt, the probability that at least one error has occurred in the preceding $m-1$ bins is:
\begin{align}
P_\text{error}\left(m\right) &= 1-\left(\frac{\plost}{1-\pdet}\right)^{m-1}.
\end{align}

The average error probability is found by summing over all possible sequences up to a total of $N$ attempts, each sequence weighted by $P\left(m^{\text{th}}~\text{click}\right)$, the probability of detecting a click on the $m^{\text{th}}$ attempt:
\begin{align}
P_\text{error} &= \sum_{m=1}^{N} P_\text{error}(m)P(m^{\text{th}}\text{~click})\nonumber\\
&= 1-\left(1-\pdet\right)^{N} - \pdet\frac{1-\plost^{N}}{1-\plost}
\label{eq:prob_error}
\end{align}
The average state fidelity after the protocol that uses sequences of length $N$ before resetting the memory is then:
\begin{align}
\mathcal{F} &= \langle\psi_{\text{ideal}}|\left(1-P_\text{error}\right)\rho_{0,\text{eff}}+P_{\text{error}}\rho_{\text{mixed}}|\psi_{\text{ideal}}\rangle
\label{eq:fidelity_total}
\end{align}
where $|\psi_{\text{ideal}}\rangle$ is the ideal transferred quantum state and $\rho_{0,\text{eff}}$ is the effective density matrix incorporating device imperfections and detection error (see Methods). We can solve Eq.~\ref{eq:fidelity_total} for the maximum number of channel uses before spin re-initialization $\Nmax$ under a given fidelity constraint, e.g. $\mathcal{F}=99\%$.

Each sequence (duration $T_{\text{seq}}$ as denoted in Fig.~\ref{fig:rate}(a)) of $\Nmax$ transmission attempts has a probability $P_\text{success} = 1-(1-\pdet)^{\Nmax}$ to detect at least one click. The number of failed sequences (i.e. each sequence of $\Nmax$ attempts without clicks followed by a memory reset) before a successful one is given by the geometric distribution. Thus, the average time of failed sequences per detector click is:
\begin{equation} 
T_{\text{failures}} = \frac{(1-\pdet)^{\Nmax}}{1-(1-\pdet)^{\Nmax}}\left[\Nmax\tpulse+\treset\right]
\label{eq:Tfail}
\end{equation}
where $4\tpulse$ is an effective pulse time accounting for repetition rate and dynamical decoupling $\pi$ pulses (see Supplemental Information).
After these failures, there is a successful sequence where the $m^{\text{th}}$ bin yields a click, which takes an average time of:
\begin{align}
T_{\text{success}} &= \treset+\sum_{m=1}^{\Nmax} P(m^{th}\text{~click})m\tpulse\nonumber\\
&= \treset+\tpulse\left(\frac{P_{\text{success}}}{\pdet}-\Nmax(1-\pdet)^{\Nmax}\right)
\label{eq:Tsucc}
\end{align}
The average quantum state transfer rate is then the inverse of the time per success:
\begin{align*}
\bar{\Gamma} = \frac{1}{T_{\text{failures}}+T_{\text{success}}}
\end{align*}

In Fig.~\ref{fig:rate}(c), we explore the trade-off between the heralded state fidelity $\mathcal{F}$ and the average rate accounting for both polarizations (see Methods). We verify our analytical solutions with Monte Carlo simulations, and show that the PEPSI can achieve transfer rate exceeding 1 kilo-qubits per second (1kqbits/sec) even at high link loss $\sim 30~$dB.

We divide the rate into three regimes. In Regime 1 (shaded red) where $\Nmax$ is low, high-fidelity state transfer prohibits increasing $\Nmax$ to offset losses in the channel, causing an exponential rate loss that intensifies for higher fidelity constraint, e.g. $\mathcal{F}=99\%$. On the other hand, for a more relaxed fidelity constraint, e.g. $\mathcal{F}=95\%$, the spin does not need to be re-initialized as frequently and the rate does not fall off as drastically.

As the link loss increases in Regime 2 (shaded blue), the number of transmission attempts per memory reset also increases. However, the time per success is still dominated by memory reset time in this regime where $\treset>\Nmax\tpulse$. As a result, the rate of increase for the number of sequences prior to detecting a click stays constant, and the rate consequently remains relatively flat.

However, in Regime 3 (shaded green) when the number of transmission attempts per sequence increases such that ${\Nmax\tpulse>\treset}$, $\etalink$ becomes the rate-limiting factor. In this Regime, the rate thus approaches the channel-limited bound (black dashed) given by $\propto\etalink/\tpulse$~\cite{Pirandola_2017}.

\section{Discussion} \label{integrated}

\begin{figure}
    \centering
    \includegraphics[width=0.5\textwidth]{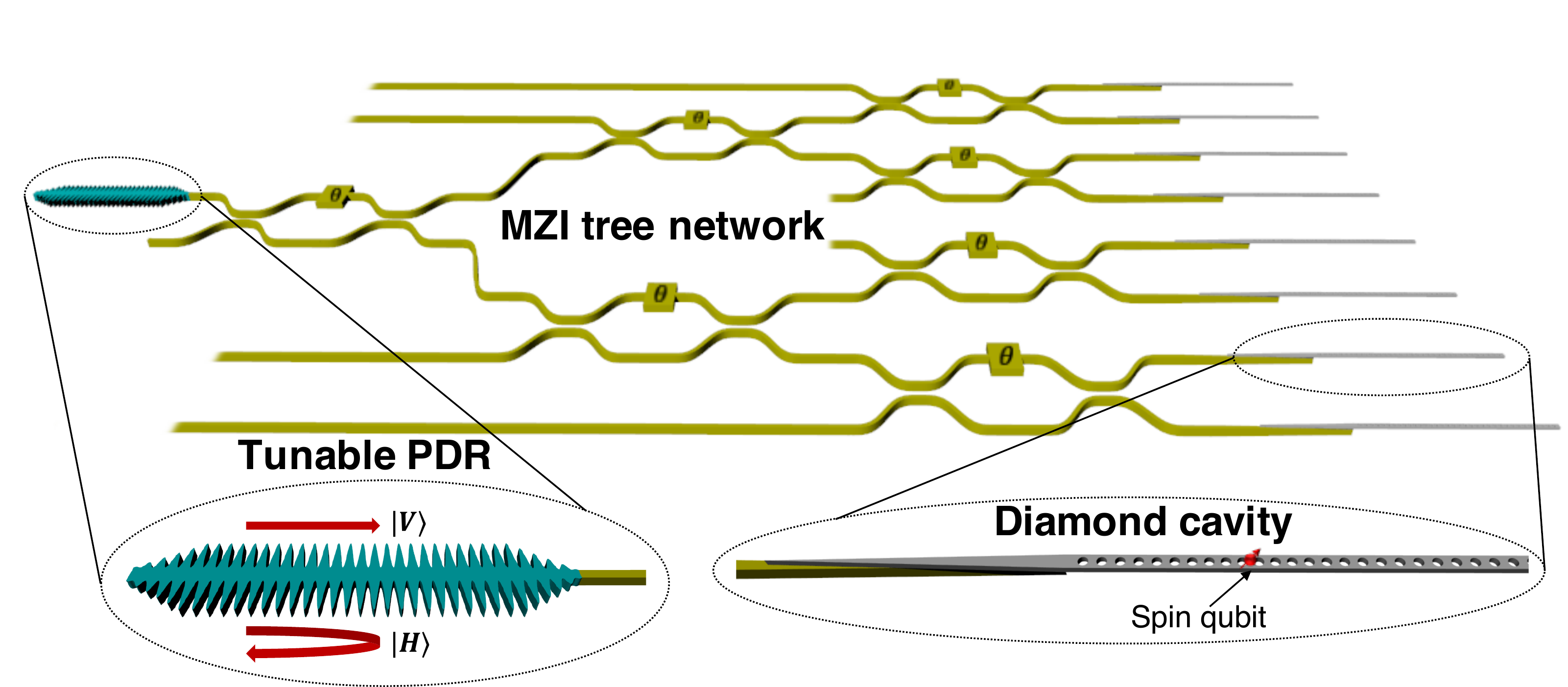}
    \caption{A PIC incorporating diamond nanocavities. The $V$ polarization passes through a tunable PDR (a combination of a PDR, active on-chip waveplates and polarizers) and enters an MZI tree network, which routes the photon to an atom-coupled cavity for quantum state transfer. Simulation (see Supplemental Information) indicates that an adiabatic taper can provide near-unity coupling efficiency between the PIC and the diamond waveguides.}
    \label{fig:integration}
\end{figure}

Practical quantum repeater nodes will likely require multiplexing over a large number of qubits. To this end, we consider the PEPSI photonic integrated circuit (PIC) illustrated in Fig.~\ref{fig:integration}. An incoming photonic qubit $\ket{\psi_P}$ enters through a PDR followed by a Mach-Zehnder interferometer (MZI) tree network, which routes the photon to a quantum memory. The PIC with $>$GHz modulation of the MZIs~\cite{Desiatov_2019} can perform state mapping across the memory array simultaneously by sending multiplexed photons to different atom-coupled cavities. As a result, the transfer rate improves by a factor of $N_{\text{cav}}$ equivalent to the number of memories connected to the tree network. The architecture can also produce heralded entanglement by sending a photon that enters an MZI 50:50 beamsplitter immediately before any two neighboring memories. Repeated heralding then produces a cluster of entangled nodes useful for quantum key distribution protocols.

Furthermore, an active PIC provides tunability essential for efficient quantum state transfer. For example, aluminum nitride photonic circuits have integrated 128 diamond waveguide-coupled color centers~\cite{Wan_2019} and can enable piezoelectric spectral tuning of photonic crystal cavities~\cite{Mouradian_2017} and diamond color center emission~\cite{Sohn_2018,Maity_2018}. Additionally, on-chip waveplates and polarizers in conjunction with the PDR (collectively termed as a tunable PDR in Fig.~\ref{fig:integration}) can perfectly balance losses to achieve high transfer fidelity. 

In summary, we introduced a phase-stable architecture for high-fidelity quantum state transfer between photonic polarization and spin qubits: the fundamental elements of a quantum repeater network. Our simulations and calculations show that the PEPSI can achieve state fidelity exceeding $99\%$ at kqbit/sec transfer rate by carefully balancing losses. Beyond color centers in diamond, our scheme applies to other quantum memories including rare-earth ions~\cite{Zhong_2015,Dibos_2018,Dutta_2020} and neutral atoms~\cite{Thompson_2013}. Additionally, we proposed a multiplexing PIC platform for state mappings across a quantum memory array via an MZI tree network. As PIC platforms have scaled beyond tens of individually controllable components~\cite{Qiang_2018}, our nanophotonic spin-photon interface should extend these gains to large-scale multiplexed quantum repeaters~\cite{Pu_2017} and even photonic cluster states~\cite{Nemoto_2014}.

\section{Methods}

\subsection{State transfer fidelity calculations}

An incoming qubit is encoded on the polarization of a photon:
\begin{equation}
\ket{\psi_P} = \alpha\ket{H}+\beta\ket{V}
\end{equation}

With the spin qubit prepared in an even superposition state $\ket{\psi_A}=\left(\ket{\downarrow}+\ket{\uparrow}\right)/\sqrt{2}$, the joint spin-photon state would be:
\begin{align}
|\psi_{\text{joint,i}}\rangle &= \alpha\ket{H,\downarrow}+\alpha\ket{H,\uparrow}+\beta\ket{V,\downarrow}+\beta\ket{V,\uparrow}
\end{align}

This photon then hits an imperfect polarization-dependent reflector (PDR) with field reflection (transmission) coefficients $r_i$ ($t_i$) for the polarization $i\in\{H,V\}$. The transmitted output is incident on a nanophotonic cavity coupled to a spin qubit. Since only the $\ket{\downarrow}\Longleftrightarrow\ket{\downarrow'}$ (see Fig.~\ref{fig:concept}(a)) transition is resonant with the cavity mode, the photon experiences a spin-dependent cavity reflection $r_{cav}\in\{r_{i,uncoupled},r_{i,coupled}\}$. The output joint photon-spin system after the HWP (that transform $H\rightarrow H+V, V\rightarrow V-H$) is in the state:
\begin{align}
\ket{\psi_f} &= \ket{H}\otimes [(\alpha r_{H,on}-\beta r_{V,on})\ket{\downarrow}\nonumber\\
\quad\quad &+(\alpha r_{H,off}-\beta r_{V,off})\ket{\uparrow}]\nonumber\\
&+\ket{V}\otimes [(\alpha r_{H,on}+\beta r_{V,on})\ket{\downarrow}\nonumber\\
\quad\quad &+(\alpha r_{H,off}+\beta r_{V,off})\ket{\uparrow}]
\end{align}
where
\begin{align}
r_{H,on} &= r_H + \frac{r_{H,coupled}t_H^2}{1-r_{H,coupled}r_H}\\
r_{H,off} &= r_H + \frac{r_{H,uncoupled}t_H^2}{1-r_{H,uncoupled}r_H}\\
r_{V,on} &= r_V + \frac{r_{V,coupled}t_V^2}{1-r_{V,coupled}r_V}\\
r_{V,off} &= r_V + \frac{r_{V,uncoupled}t_V^2}{1-r_{V,uncoupled}r_V}
\end{align}
Note that the transmissivity is modified with the insertion of a polarizer: $|t_{V/H}|^2 \rightarrow \eta_{\text{pol,V/H}}|t_{V/H}|^2$. A detailed derivation is presented in the Supplementary Information.

The state-dependent cavity reflection coefficients can be derived from using the input-output formalism~\cite{Waks_2006_PRA,Reiserer_2015}:
\begin{align}
r(\omega)&= 1-\frac{\kappa_{wg}}{i(\omega_c-\omega)+\frac{\kappa}{2}}\frac{1}{1+\frac{g^2}{\left(i(\omega_c-\omega)+\frac{\kappa}{2}\right)\left(i(\omega_a-\omega)+\frac{\gamma}{2}\right)}}\\
&= \frac{C-1}{C+1} \quad (\text{large} \ C \ \text{limit})
\end{align}
where $\kappa,\gamma,g$ are the cavity total decay, atom relaxation, and atom-cavity coupling rates. $\kappa_{wg}$ is the cavity decay rate into the waveguide. $(\omega_c-\omega)$ and $(\omega_a-\omega)$ are the cavity and atom detuning, respectively. $C=\frac{4g^2}{\kappa\gamma}$ is the atom-cavity cooperativity. In the uncoupled case, $g=0$ and a bare reflection off the cavity gives the photon a $-1$ phase. On the other hand, the coupled state results in a $+1$ phase. The relative phase conditioned on the atomic state forms the basis behind the state transfer protocol detailed in Ref.~\cite{Duan_2004}.

Finally, a detection of an $\ket{D}$ or $\ket{A}$ photon heralds mapping of the input photonic state onto the spin with an additional Hadamard rotation on the spin (and a conditional $\pi$ rotation). We can calculate the state fidelity by Eq.~\ref{eq:fid_single}.

\subsection{Transfer rate calculations}

In the state transfer rate calculations, we consider the detection, scattering loss, and error paths for both polarizations and compute the average rate. Here we denote the power transmission and reflection coefficients of the PDR as $T_{V/H}$ and $R_{V/H}$. The cavity reflectivity $\rcavV/\rcavH$ is the average reflectivity between on- and off-resonance cases for the $V/H$ polarization, respectively: $\rcavV=(|r_{\text{cavity-V}}|^2+|r_{\text{coup-V}}|^2)/2=35.6\%, \rcavH=92.1\%$ basing on our simulated device and $C=4$. The $V$-pass polarizer has $V$ and $H$ transmission efficiencies to be $98.9~\%$ and $12.8~\%$, respectively.

In order to detect a $V$ ($H$) photon, it has to either undergo a roundtrip as denoted by Fig.~\ref{fig:rate}(b) with probability $\etalink T_V\etapol\rcavV\etapolV T_V\etadet$ ($\etalink T_H\etapol\rcavH\etapolH T_H\etadet$) or reflect off the PDR upon the first pass with probability $\etalink R_V\etadet$ ($\etalink R_H\etadet$), where $\etadet=93.6\%$ is the detection efficiency accounting for the PBS, the HWP, and the photon detector (see Supplementary Information). The average detection probability is then
\begin{align}
\pdet &= \frac{\etalink}{2}\left(T_V^2\etapolV^2\rcavV+R_V\right.\nonumber\\
&+\left.T_H^2\etapolH^2\rcavH+R_H\right)\etadet
\end{align}

The photon can also scatter off en route to the PDR with probability $1-\etalink$, by the PDR, or by the polarizer, contributing to the probability of photon loss \textit{without} erroring:
\begin{align}
\plost &= 1-\etalink+\frac{\etalink}{2}\left(\zeta_V+\zeta_H+T_V(1-\etapolV)\right.\nonumber\\
&+\left.T_H(1-\etapolH)+T_H\etapolH(1-\rcavH-\xi)\right)
\end{align}
where $\zeta_V=1-T_V-R_V$ and $\zeta_H=1-T_H-R_H$. $\xi$ is the probability of the $H$ photon reaching the cavity where the spin qubit resides (see Supplemental Information).

Lastly, as addressed in Sec.~\ref{sec_rate}, the photon can be lost after interacting with the atom-cavity system, hence yielding an unheralded error. Specifically, we consider errors arising from the $V$ photon scattering after cavity interaction and the small amount of $H$ photon leaking into the cavity:
\begin{align}
p_e &= 1-\pdet-\plost\\
&= \frac{\etalink}{2}\left[T_V\etapolV(1-\rcavV+\rcavV(1-\etapolV)\right.\nonumber\\
&\left.+\rcavV\etapolV\zeta_V)+T_H\etapolH\xi\right]
\end{align}

The probabilities are used to compute $P_{\text{error}}$, which is the probability of at least one unheralded error occurring after detecting the first and only click on the $m^{\text{th}}$ attempt within $N$ attempts. After which, we can calculate the average state fidelity:
\begin{align}
\mathcal{F} &= \langle\psi_{\text{ideal}}|\left(1-P_\text{error}\right)\rho_0+P_{\text{error}}\rho_{\text{mixed}}|\psi_{\text{ideal}}\rangle
\end{align}
where $\rho_0$ is the density matrix corresponding to a single attempt considering only device imperfections.

\subsection{Monte Carlo simulations}

Numerical simulations were performed in MATLAB (MathWorks Inc.). For each attempt, a probability value $p_{\text{random}}$ is chosen out of a uniform distribution $U(0,1)$. $p_{\text{random}}$ then determines if the photon is lost in the device before reaching the spin ($\plost$), lost in the device \textit{after} the spin ($p_e$), or detected ($\pdet$). Each simulation trial terminates once the photon is detected, and the total experimental time is recorded. A trial can consist of multiple sequences, and each sequence has the number of attempts up to $\Nmax$, which depends on the fidelity constraint and the link loss. Each Monte Carlo data point presented in Fig.~\ref{fig:rate}(c) is the average rate of 100 simulation trials.

\section{Acknowledgements}

K.C.C. acknowledges funding support by the National Science Foundation Graduate Research Fellowships Program (GRFP) and the Army  Research  Laboratory  Center  for  Distributed Quantum Information (CDQI). E.B. is supported by a NASA Space Technology Research Fellowship and the NSF Center for Ultracold Atoms (CUA). D.E. acknowledges support from the NSF EFRI-ACQUIRE program Scalable Quantum Communications with Error-Corrected Semiconductor Qubits. We also thank Ian Christen for valuable discussions.

\section{Author contributions}

K.C.C. performed the FDTD simulations in designing the PEPSI. K.C.C. and E.B. calculated the average state fidelity and the transfer rate. D.E. conceived the idea. All authors contributed to writing and revising the manuscript.

\bibliography{references}

\end{document}